\documentclass[twocolumn,prl,preprintnumbers,amsmath,amssymb,superscriptaddress,aps]{revtex4}
\pdfoutput=1
\usepackage[font=footnotesize,labelfont=sf,bf,flushleft]{caption}
\usepackage{graphicx}
\usepackage[usenames,dvipsnames]{xcolor}
\usepackage{amsmath,bbm}

\begin{document}

\title{Optically driven ultra-stable nanomechanical rotor}

\author{Stefan Kuhn}
\email{stefan.kuhn@univie.ac.at}
\affiliation{University of Vienna, Faculty of Physics, VCQ, Boltzmanngasse 5, 1090 Vienna, Austria}
\author{Benjamin A. Stickler}
\affiliation{University of Duisburg-Essen, Lotharstra{\ss}e 1, 47048 Duisburg, Germany}

\author{Alon Kosloff}
\affiliation{School of Chemistry, Tel-Aviv University, Ramat-Aviv 69978, Israel}
\author{Fernando Patolsky}
\affiliation{School of Chemistry, Tel-Aviv University, Ramat-Aviv 69978, Israel}
\author{Klaus Hornberger}
\affiliation{University of Duisburg-Essen, Lotharstra{\ss}e 1, 47048 Duisburg, Germany}
\author{Markus Arndt}
 \affiliation{University of Vienna, Faculty of Physics, VCQ, Boltzmanngasse 5, 1090 Vienna, Austria}
\author{James Millen}
 \affiliation{University of Vienna, Faculty of Physics, VCQ, Boltzmanngasse 5, 1090 Vienna, Austria}

\maketitle


\noindent \textbf{Abstract}

\noindent \textbf{Nanomechanical devices have attracted the interest of a growing interdisciplinary research community, since they can be used as highly sensitive transducers for various physical quantities. Exquisite control over these systems facilitates experiments on the foundations of physics. Here, we demonstrate that an optically trapped silicon nanorod, set into rotation at MHz frequencies, can be locked to an external clock, transducing the properties of the time standard to the rod's motion with the remarkable frequency stability $f_{\rm r}/\Delta f_{\rm r}$ of $7.7 \times 10^{11}$. While the dynamics of this periodically driven rotor generally can be chaotic, we derive and verify that stable limit cycles exist over a surprisingly wide parameter range. This robustness should enable, in principle, measurements of external torques with sensitivities better than 0.25\,zNm, even at room temperature. We show that in a dilute gas, real-time phase measurements on the locked nanorod transduce pressure values with a sensitivity of 0.3\,\%.}\\
%

\begin{figure}[!ht]
	 {\includegraphics[width=0.46\textwidth]{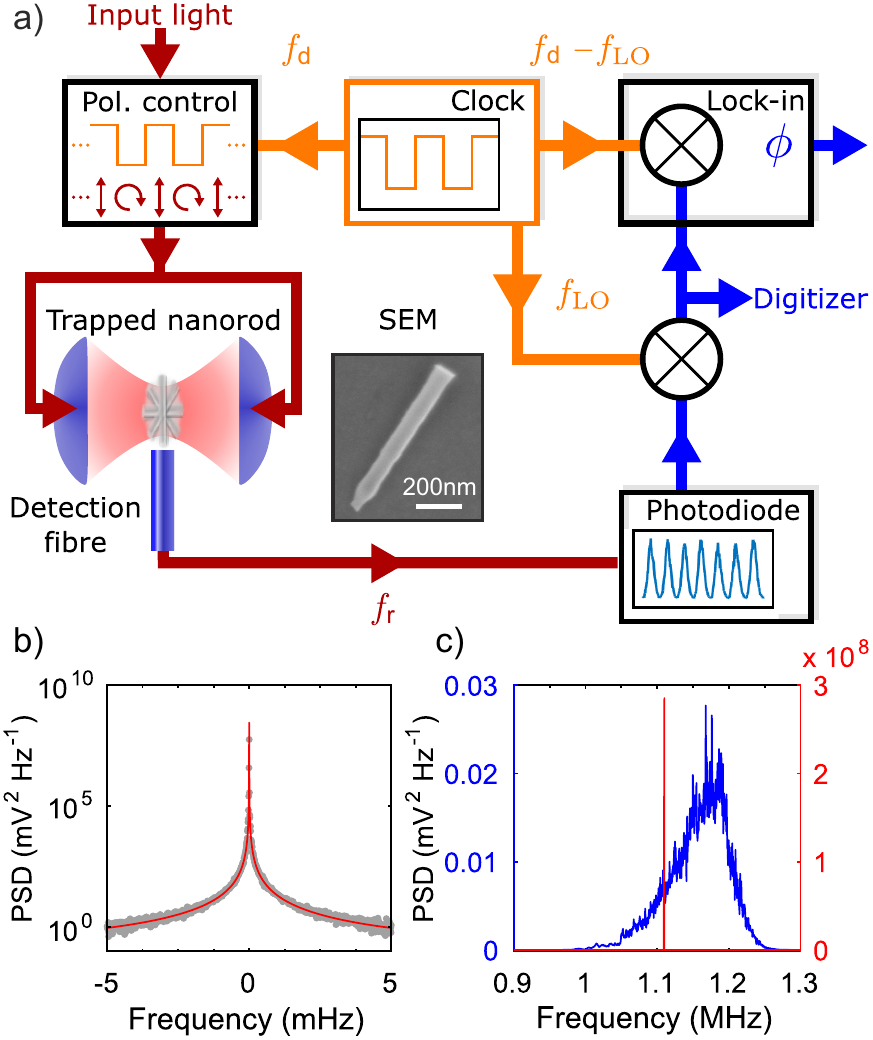}}	
\caption{\label{fig:layout} 
Frequency locking. a) A silicon nanorod is optically trapped in a standing-wave formed by counterpropagating focussed laser beams at $\lambda = 1550\,$nm, see the Methods section. The polarization of the light is controlled using a fibre-EOM driven by a signal generator. We detect the motion of the nanorod via the scattered light collected in a multimode optical fibre. The signal is mixed down with a local oscillator $f_{\rm LO}$ to record the spectrum and the phase of the rotation with respect to the drive frequency $f_{\rm d}$. Both frequencies $f_{\rm d}$ and $f_{\rm LO}$ are synced to a common clock. b) Power spectral density (gray points) of the frequency-locked rotation at 1.11\,MHz, taken over 4 continuous days, fit with a Lorentzian (red curve).
The upper bound on the FWHM is 1.3\,$\mu$Hz. c) Comparison of driven rotation when frequency-locked (``locked rotation'', red, right axis) and un-locked (``threshold rotation'', blue, left axis).}
\end{figure}

\noindent \textbf{Introduction}\\
Frequency is the most precisely measured quantity in physics, and stable oscillators have found a plethora of applications in metrology. While pendulum clocks exploited the stability of mechanical motion to keep track of time, state-of-the-art atomic clocks rely on well-defined internal resonances of atoms, achieving a precision of a few parts in $10^{18}$ \cite{Ludlow2015}. To exploit the stability of clocks for physical applications, it is essential to develop ``gearboxes'' that can be synchronized to a good time standard, translating atomic definiteness into other domains of physics. Phase-locked quartz oscillators \cite{Vig1999} and frequency combs are examples of such transducers, imprinting clock stability onto a mechanical system or light field respectively, with high accuracy \cite{Udem2002}. 

Nano- and micromechanical systems are of great technological interest, due to their low mass and extreme sensitivity to external forces \cite{Rugar01, Kippenberg12, Geraci15}, torques \cite{Kim13, Kim16}, acceleration \cite{Painter12}, displacement \cite{Lehnert09, Wilson15}, charge \cite{Roukes98}, and added mass \cite{Bachtold12, Zadeh13}. Many of these systems are themselves realizations of harmonic oscillators, with frequency stabilities reaching $f/ \Delta f = 10^8$ \cite{Norte16, Schliesser16}, which can be further improved through mechanical engineering \cite{Ghaffari14}, injection locking \cite{Vahala08, Coppock16}, electronic feedback \cite{Roukes08}, or parametric driving \cite{Villanueva11}. The contact-free mechanical motion of particles suspended by external fields in vacuum \cite{Twamley12, Gieseler12, Yin13, Kiesel13, Vamivakas14, Millen15, Hoang2016, Kuhn16}, can reach a frequency stability that is only limited by laser power fluctuations and collisional damping by residual gas particles \cite{Chang10}. 

In this work, we transduce clock stability into the rotation of an optically trapped silicon nanorod in vacuum. By periodically driving the rotation with circularly polarized light, we create a nanomechanical rotor whose rotation frequency $f_\mathrm{r}$, and frequency noise, is determined by the periodic drive alone. This driven rotor is sensitive to non-conservative forces, and the operating frequency can be tuned by almost $10^{12}$ times its linewidth. Through our method, the frequency stability is independent of material stress, laser noise, and collisional damping. The driven nanorotor operates at room temperature, and across a wide pressure range from low vacuum to medium vacuum, achieving a pressure resolution of 3 parts-per-thousand, and in principle allowing a torque sensitivity below the zepto-Nm level.\\

\noindent {\bf Results}\\
\noindent  {\bf Frequency locking}\\ 
We levitate a nanofabricated silicon nanorod in the standing light wave formed by two counterpropagating laser beams, and track its motion by monitoring the scattered light, see Fig.\ref{fig:layout}a and Methods. When the laser is linearly polarized, the nanorod is harmonically trapped in an antinode of the standing wave and aligned with the field polarization. When the laser is circularly polarized, the scattered light exerts a torque \cite{Tong09, Dholakia13, Kuhn16} and propels the nanorod in the plane orthogonal to the beam axis, while its centre-of-mass remains trapped \cite{Kuhn16}. The maximum rotation frequency of the rod is determined by its size, the pressure, and the laser intensity \cite{Kuhn16}. Collisions with gas particles, and centre-of-mass excursions into regions of different light intensity, give rise to a broad distribution of rotational frequencies \cite{Kuhn16}, as shown by the blue curve in Fig.~\ref{fig:layout}c.

However, if the rod is driven by periodically switching the laser polarization between linear and circular, the rod can frequency-lock to this modulation, leading to a sharp peak in the power spectral density (PSD) of the scattered light (Fig.~\ref{fig:layout}b). The locked rotation peak is eleven orders of magnitude narrower than in the unlocked case (see Fig.\ref{fig:layout}c). The rotation frequency can be continuously tuned over a range of $10^{12}$ linewidths whilst locked to the periodic drive, retaining its high frequency stability.

\begin{figure}
	 {\includegraphics[width=0.4\textwidth]{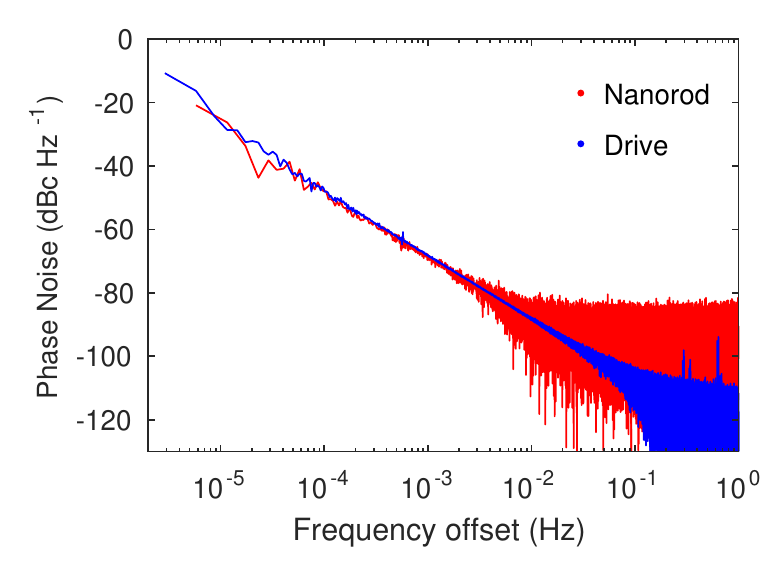}}	
\caption{\label{fig:freq} 
Characterizing the locked rotation. The phase noise of the rotor (red) and drive (blue) is calculated from two individual four day measurements at a drive frequency of $f_{\rm d} = 1.11$\,MHz (see Methods).
}
\end{figure}

In order to characterize the frequency-locked rotation, we drive the nanorod with frequency $f_{\mathrm{d}} = 1.11\,$MHz, at a gas pressure $p_{\rm g} = 4\,$mbar and total laser power $P = 1.35\,$W. We record its motion continuously for four days (see Methods section). The PSD of the locked rotation is shown in Fig.~\ref{fig:layout}b. We record an extremely narrow feature, which yields a Lorentzian FWHM below $1.3\,\mu$Hz within one standard deviation. In this way, we achieve rotational stability with a phase noise $-80\,$dBc\,Hz$^{-1}$ below the signal at only $3$\,mHz from the carrier frequency (Fig.~\ref{fig:freq}).\\

\noindent {\bf Stable limit cycles}\\ 
To explain this ultra-stable rotation, we model the dynamics of the rod's orientation $\alpha$ with respect to the polarization axis \cite{Kuhn16}. Denoting by $\Gamma$ the damping rate due to gas collisions (a function of gas pressure $p_{\rm g}$), by $N$ the torque exerted by the circularly polarized standing wave (a function of laser power $P$) and by $V$ the maximum potential energy of the rod in a linearly polarized standing wave, the equation of motion is

\begin{align}
\label{eqn:eom}
\ddot{\alpha} = -\Gamma\dot{\alpha} + \frac{N}{I}h(t) - \frac{V}{I}\sin(2\alpha)[1 - h(t)],
\end{align}

\noindent where $I = M \ell^2/12$ is the moment of inertia and $h(t)$ represents the periodic driving, with $h(t) = 1$ for circular polarization at $t \in [0, 1/2 f_{\rm d})$ and $h(t) = 0$ for linear polarization at $t \in [1/2 f_{\rm d}, 1/f_{\rm d})$; expressions for $\Gamma$, $N$, and $V$ are given in the Methods section.

In the limit of long driving times $t \gg 1/f_{\rm d}$ the nanorod rotates with constant mean rotation frequency $f_{\rm r} = \langle \dot{\alpha} \rangle/2 \pi$, and its motion approaches one of two qualitatively distinct types of limit cycle. In the first type, threshold rotation, $f_{\rm r} = N/(4 \pi I\Gamma)$ is determined by the balance between $\Gamma$ and the time averaged radiation torque $N/2$. Threshold rotation exhibits a broad frequency distribution, as shown by the blue curve in Fig.~\ref{fig:layout}c, due to its dependence on $P$ and $p_{\rm g}$ \cite{Kuhn16}. It should be noted, that the behaviour of threshold rotation is almost identical to illuminating the nanorotor continuously with circularly polarized light, which propels the rod at a maximum rotation frequency $f_{\rm r} = N/(2 \pi I\Gamma)$.

In the second type of limit cycle, the aforementioned locked rotation, $f_{\rm r}$ locks to a rational fraction of the driving frequency. Experimentally, we observe $f_{\rm r}:f_{\rm d} = 1:2$~- and $f_{\rm r}:f_{\rm d} = 1:4$~-~locking, where the rod performs one rotation in two or four driving periods, respectively. The rotational frequency $f_{\rm r}$ now does not depend on environmental parameters such as $p_{\rm g}$ or $P$, but only on the frequency stability of the drive. Locked rotation is shown in Fig.~\ref{fig:layout}b and the red curve in Fig.~\ref{fig:layout}c.

\begin{figure}
	 {\includegraphics[width=0.46\textwidth]{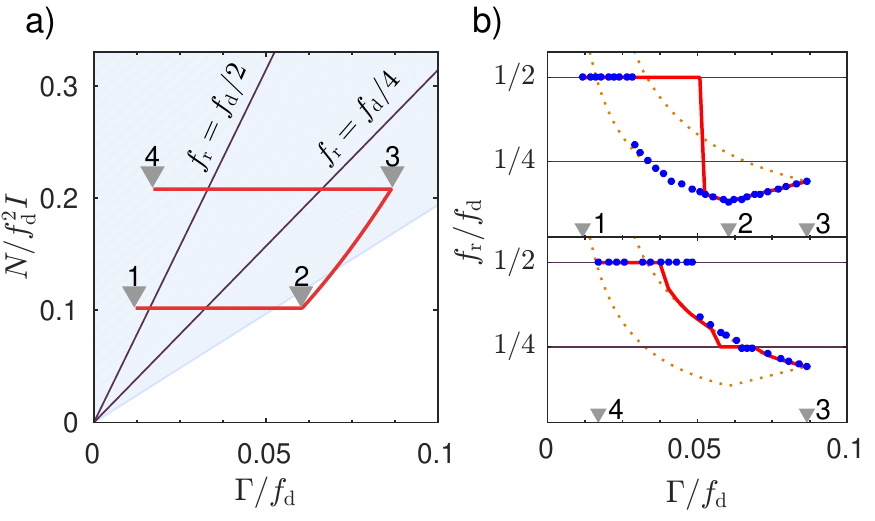}}	
\caption{\label{fig:limitcycle} 
Driven rotor. a) Reduced parameter space for the periodically driven nanorod. Different regions can exhibit different limit cycles, as determined by the initial conditions. Within the blue shaded region $f_{\rm r}:f_{\rm d} = 1:2$~-~locking can occur, while it can not be realized in the white region. Along the labelled solid lines the frequency of threshold rotation coincides with $1:2$~- or $1:4$~-~locking. To explore the various limit cycles we select a path in parameter space (solid red line) which we follow experimentally by adiabatically varying $p_{\rm g}$ and $f_{\rm d}$. b) Experimentally measured rotation frequencies $f_{\rm r}$ along the path in parameter space (blue points), and simulation (solid red line). The three different limit cycles observed are: threshold rotation (orange dotted lines), $1:2$~- and $1:4$~-~locking (horizontal solid lines).}
\end{figure}

The realized limit cycle is determined by the initial dynamics of the nanorod, and by experimental parameters such as $p_{\rm g}, P, f_{\rm d}$, and the rod dimensions. For a given nanorod the ratio between torque and potential is fixed, and thus Eq.~\eqref{eqn:eom} depends only on the dimensionless damping rate $\Gamma / f_{\rm d}$ and dimensionless torque $N / f_{\rm d}^2 I$. In Fig. \ref{fig:limitcycle}a we show this reduced parameter space. The blue shaded area indicates the region where $1:2$~-~locking is possible. The labelled solid lines indicate where threshold and locked rotations have the same frequency, and locking occurs independent of the initial conditions. \\
\\  

\noindent {\bf Rotational dynamics}\\ 
To explore the dynamics of the driven nanorotor, we experimentally vary $p_{\rm g}$ and $f_{\rm d}$, thereby following a path through parameter space (red line in Fig.~\ref{fig:limitcycle}a). The observed $f_{\rm r}$ are shown as blue points in Fig.~\ref{fig:limitcycle}b, the top panel showing the path 1-2-3, and the lower panel 3-4. This plot shows that for certain parameters both types of limit cycles can be observed, depending on the initial conditions. When starting from 1, the rotation $1:2$~-~locks to the drive (horizontal solid line). When increasing $p_{\rm g}$ along 1-2, the rod jumps out of lock, and exhibits threshold rotation. When decreasing $f_{\rm d}$ along 2-3, the rod remains at threshold rotation, following the theoretically expected frequencies with excellent agreement (orange dotted lines). Decreasing $p_{\rm g}$ along 3-4, the rod first follows the threshold rotation frequency until it crosses the horizontal line, where it briefly enters $1:4$~-~locked rotation, returns to threshold rotation and eventually jumps into $1:2$~-~locked rotation. The solid red lines in Fig.~\ref{fig:limitcycle}b are the theoretically predicted $f_{\rm r}$ for an adiabatic path through parameter space, with the discrepancy due to imperfect adiabatic control in the experiment. \\
\\

\noindent {\bf Applications}\\
For $1:2$~-~locked rotation, the phase lag $\phi$ between the drive and the rotation is

\begin{align}
\label{eqn:phase}
\phi = \arccos \Bigg[ \frac{\pi}{2V} \Bigg( N - 2\pi f_{\rm d} I \Gamma \Bigg)\Bigg];
\end{align}

\noindent it depends upon gas pressure $p_{\rm g}$ through $\Gamma$ and on laser power $P$ through $V$ and $N$. The requirement that this phase is real defines the shaded region in Fig.~\ref{fig:limitcycle}a. The phase $\phi$ is sensitive to non-conservative forces, such as light or gas scattering.

\begin{figure}[t]
	 {\includegraphics[width=0.46\textwidth]{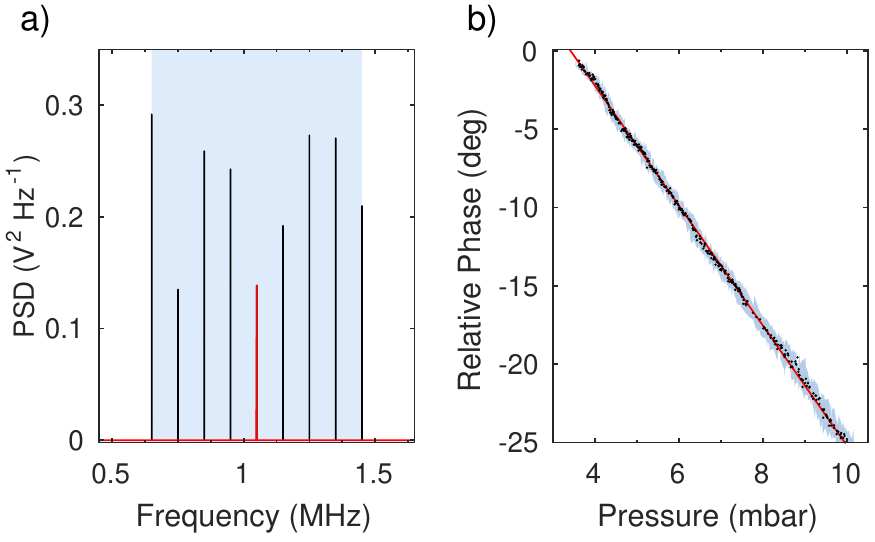}}	
\caption{\label{fig:application} 
Sensing applications. a) At 4\,mbar, the locking frequency $f_{\rm d}$ can be continuously tuned by over 800\,kHz, which corresponds to almost $10^{12}$ linewidths. The peaks in the PSD differ in amplitude since for short recording times they are not fully resolved. b) In this parameter region the phase lag (black dots) between the rotor and the drive depends linearly on pressure (fitted red line) and thus can be used for precise pressure sensing. The pressure values are given by the reading of a commercial pressure gauge (Pfeiffer Vacuum PCR 280), and the shaded region is the error margin as defined by the manufacturer-given resolution and repeatability of the gauge.
}
\end{figure}

A lock-in amplifier is used to monitor $\phi$ (see the Methods section), yielding real-time readout of phase variations. For a constant laser power $P$, measuring the phase amounts to local sensing of the gas pressure $p_{\rm g}$, as shown in Fig.~\ref{fig:application}b. We achieve a relative pressure sensitivity of $0.3$\,\%, which is currently limited only by the intensity noise of the fiber amplifier. This sensitivity may still be improved by five to six orders of magnitude by stabilizing the power \cite{Kwee11}. This shows the great potential of our system as a pressure sensor. The fine spatial resolution provided by the micrometer-sized rotor could allow, for instance, mapping of velocity fields and turbulences in rarefied gas flows or atomic beams.

The driven nanorod is sensitive to externally applied torques through Eq.~\eqref{eqn:eom}. In analogy to the pressure sensing application, the phase between the drive and the driven rotation of the locked rotor provides a real-time readout with a bandwidth of 100\,kHz set by the lock-in amplifier. From Eq.~\eqref{eqn:phase}, the torque sensitivity can be estimated as $2.4 \times 10^{-22}$\,Nm for our current experimental parameters, which at room temperature would be the highest value achieved in state of the art systems~\cite{Kim16}, and which could also be significantly improved by laser power stabilization. This sensitivity can be reached for any non-conservative force, that varies on a timescale longer than $1/f_{\rm d}$. The bandwidth and sensitivity of this sensor for arbitrary torques warrants further investigation. \\

\noindent {\bf Discussion} \\
In conclusion, we have presented a technique for locking the rotation of a levitated nanomechanical rotor to a stable frequency reference, leading to a high mechanical frequency stability which can be used for instantaneous, highly sensitive local measurements of pressure or torque. One great advantage of the locked nanorotor over nanomechanical resonators is the absence of an intrinsic resonance frequency. By varying $f_{\rm d}$ we can tune the rotation frequency over almost $10^{12}$ linewidths while retaining its stability, as shown in Fig.~\ref{fig:application}a, circumventing the low bandwidth that naturally comes with highly sensitive resonant detectors. Measuring a phase lag rather than a frequency shift allows us to monitor force variations in real time, bypassing the inherent growth in measurement time that comes with decreasing linewidth in a resonant sensor. 

Our system is sensitive to non-conservative forces, such as those due to photon absorption or emission, shear forces in gas flows, radiation pressures in light fields, optical potentials, mass and size variations of the rod due to gas accommodation, and local pressure and temperature variations in the gas. Employing higher optical powers, larger duty cycles of the drive, or lower gas pressures will increase the sensing range by a factor of more than ten, while retaining the exquisite sensitivity. The stability of the locked nanorotor may be further increased by driving it with a more stable clock. By reaching ultra high vacuum, this technique may also be suited to prepare quantum coherent rotational dynamics \cite{Bhattacharya16}, for which the high frequency stability may be exploited.

\vspace{5mm}

\noindent \textbf{Methods}\\

\noindent {\bf Nanorod trapping}\\
A silicon nanorod of length $\ell = (725\pm15)\,$nm and diameter $d = (130\pm13)\,$nm (with mass $M = 2.2 \times 10^{-17}$\,kg) is trapped at a pressure of $p_{\rm g} = 4\,$mbar, using light of total power $P = 1.35$\,W with RMS power fluctuations of 0.3\,\%. Our method of producing and trapping the nanorods is described in Refs. \cite{Kuhn15, Kuhn16}. The motion is monitored by placing a 1\,mm core multimode fibre less than $100\,\mu$m from the trapped nanorod, which collects the light that the nanoparticle scatters, yielding information about all translational and rotational degrees of freedom.

\noindent {\bf Rotation analysis}\\ 
In order to record a time series as long as 4 days we mix down the scattered light with a local oscillator at frequency $f_{\rm LO}$ such that the rotational motion signal of the rod is shifted to a frequency of 190\,Hz (see Fig.~\ref{fig:layout}a) and digitized with a sampling rate of $2$\,kS\,s$^{-1}$. This signal can then be used to calculate the phase noise $S_{\phi}(f) = 10 \log_{10}\left[{\rm PSD}(f)/{\rm PSD}(f_{\rm d})\right]$ in units of dBc\,Hz$^{-1}$. The drive signal is recorded and analysed in the same way.

To extract the relative phase $\phi$ of the nanorod rotation with respect to the drive we use a Stanford Research Systems lock-in amplifier (SR830) which performs a homodyne measurement on the mixed-down 190\,Hz scattering signal. For this purpose both the modulation frequency $f_{\rm d}$ and the local oscillator $f_{\rm LO}$ are synced to a common clock. 

\noindent {\bf Rotational dynamics}\\ 
The rotational motion depends crucially on the damping rate $\Gamma$, the radiation torque $N$, and the laser potential $V$, through Eq.~\eqref{eqn:eom}. All three quantities can be evaluated microscopically as detailed in Refs. \cite{Kuhn16,Stickler16}. Specifically, the rotational damping rate due to diffuse reflection of gas atoms with mass $m_{\rm g}$, evaluated in the free molecular regime, is $\Gamma = d \ell p_{\rm g} \sqrt{2 \pi m_{\rm g}}(6 + \pi)/8 M \sqrt{k_{\rm B} T}$, where $T$ denotes the gas temperature. The optical torque exerted by a circularly polarized standing wave can be evaluated by approximating the internal polarization field according to the generalized Rayleigh-Gans approximation \cite{Stickler16} as $N = P \Delta \chi \ell^2 d^4 k^3 [ \Delta \chi \eta_1(k \ell) + \chi_\bot \eta_2(k \ell) ] / 48 c w_0^2$, where $k = 2\pi / \lambda$ is the wavenumber, $\Delta \chi = \chi_\| - \chi_\bot$ depends on the two independent components of the susceptibility tensor and $\eta_1(k \ell) = 0.872$ and $\eta_2(k \ell) = 0.113$ \cite{Kuhn16}. In a similar fashion, one obtains for the laser potential $V = P d^2 \ell \Delta \chi/ 2 c w_0^2$.

The phase lag~\eqref{eqn:phase} is obtained by averaging the equation of motion~\eqref{eqn:eom} over one driving period $1/f_{\rm d}$ and exploiting that the motion is 1:2-locked, $\langle \dot{\alpha}\rangle = \pi f_{\rm d}$ while $\langle \ddot{\alpha}\rangle = 0$ and, consequently, $\alpha(t) = \alpha_0 + \pi f_{\rm d} t$. The phase difference between the polarization change from circular to linear and the maximum scattering signal observed when the rod is oriented orthogonal to the detector ($\alpha(t) = \pi$) is $\phi = \pi - 2 \alpha_0$ leading to Eq.~\eqref{eqn:phase}.\\

\noindent \textbf{Data Availability}\\
All relevant data are available from the corresponding author upon request.

\vspace{5mm}

\noindent \textbf{Acknowledgments} 	\\
\noindent We are grateful for financial support by the Austrian Science Fund (FWF) in the projects P27297 and DK-CoQuS (W1210). We acknowledge support by S. Puchegger and the faculty center for nanostructure research at the University of Vienna in imaging the nanorods. JM acknowledges funding from the European Union’s Horizon 2020 research and innovation programme under the Marie Sk{\l}odowska-Curie grant agreement No 654532. F.P. acknowledges the Legacy Program (Israel Science Foundation) for its support.

\vspace{5mm}

\noindent \textbf{Author Contributions} \\
\noindent S.K., M.A. and J.M. conceived of and designed the experiment. S.K. and J.M. performed all experiments. S.K., B.A.S. and J.M. analysed the data. A.K. and F.P. fabricated the nanorods. B.A.S. and K.H. developed all theoretical models. S.K., B.A.S., K.H., M.A. and J.M. contributed to the writing of the manuscript.

\vspace{5mm}

\noindent \textbf{Competing Financial Interests} \\
\noindent Aspects of this work are subject to a filed patent application.

\end{document}